\documentclass[11pt,preprint]{aastex}
\newcommand{\be}{\begin{equation}}
\newcommand{\ba}{\begin{eqnarray}}
\newcommand{\ee}{\end{equation}}
\newcommand{\ea}{\end{eqnarray}}
\newcommand{\etal}{et al.\ }
\newcommand{\etalb}{et al.}

\newcommand{\Ni}{N_{\rm ion}}
\newcommand{\Omm}{\Omega_m}

\newcommand{\cN}{c_{\rm N}}
\newcommand{\Lya}{\mbox{Ly$\alpha$ }}

\begin{document}
\title{Effective Screening due to Minihalos During the Epoch of Reionization}

\author{Rennan Barkana}
\affil{School of Physics and Astronomy, Tel Aviv University, Tel Aviv 69978,
ISRAEL}
\email{barkana@wise.tau.ac.il}

\author{Abraham Loeb} 
\affil{Department of Astronomy, Harvard University, 60 Garden St.,
Cambridge, MA 02138}
\email{aloeb@cfa.harvard.edu}

\begin{abstract}
We show that the gaseous halos of collapsed objects introduce a
substantial cumulative opacity to ionizing radiation, even after the
smoothly distributed hydrogen in the intergalactic medium has been
fully reionized. This opacity causes a delay of $\Delta z \ga 1$ in
redshift between the time of the overlap of ionized bubbles in the
intergalactic medium and the lifting of complete Gunn-Peterson \Lya
absorption. The minihalos responsible for this screening effect are
not resolved by existing numerical simulations of reionization.
\end{abstract}



\section{Introduction}
\label{sec:Intro}

The spectra of the quasars SDSS 1030+0524 at $z=6.28$
\citep{z6.3} and SDSS 1044-0125 at $z=5.73$ \citep{z5.8} 
feature broad regions of high optical depth, possibly implying the
existence of a Gunn-Peterson trough of complete resonant Ly$\alpha$
absorption by a neutral intergalactic medium \citep{GP}. However, the
discovery of a Ly$\alpha$ emission line from a galaxy at $z=6.56$
\citep{z6.6} implies that the gas in the vicinity of that galaxy is
mostly ionized and that the neutral fraction is highly
non-uniform. The appearance of a Gunn-Peterson trough is generally
attributed to absorption by a smooth intergalactic medium (hereafter
IGM) with a significant neutral fraction, and it indicates that
observations are approaching the reionization era.

The reionization of hydrogen is expected to involve several distinct
stages. The initial, ``pre-overlap'' stage [using the terminology of
\citet{G00}] consists of individual ionizing sources turning on and
ionizing their surroundings. The radiation from the first galaxies
must make its way through the surrounding gas inside the host halo,
then through the high-density region which typically surrounds each
halo. Once they emerge, the ionization fronts propagate more easily
into the low-density voids, leaving behind pockets of neutral,
high-density gas. During this period the IGM is a two-phase medium
characterized by highly-ionized regions separated from neutral regions
by sharp ionization fronts. The main, relatively rapid ``overlap''
phase of reionization begins when neighboring \ion{H}{2} regions start
to overlap. Whenever two ionized bubbles are joined, each point inside
their common boundary becomes exposed to ionizing photons from both
sources. Therefore, the ionizing intensity inside \ion{H}{2} regions
rises rapidly, allowing those regions to expand into high-density
gas. This process leads to a state in which the low-density IGM is
highly ionized and ionizing radiation reaches everywhere except for
gas located inside self-shielded, high-density clouds. The ionizing
intensity continues to grow during the ``post-overlap'' phase, as an
increasing number of ionizing sources becomes visible to every point
in the IGM.

The recent observations of strong \Lya absorption at $z \sim 6$
indicate a significant neutral fraction in the IGM
\citep[e.g.,][]{Fan02,Lidz}. However, they do not necessarily
correspond to the overlap stage of reionization, since even after the
end of overlap, variations in the gas density and the ionizing
intensity in the IGM may be able to produce such large regions of
absorption \citep{z6me}. When the voids become transparent to ionizing
photons, the naive expectation is for the ionizing intensity to
increase quickly as multiple ionizing sources begin to reach every
point in the IGM. However, such a sudden, homogeneous end to the
Gunn-Peterson trough may conflict with the observations, which imply
that along one line of sight, overlap occurs at $z > 5.9$
\citep{z6.3}, while a relatively large region of strong absorption is
found along a second line of sight at $z=5.3$ \citep{z5.8};
furthermore, the recent $z=6.56$ galaxy may indicate the existence of
a transparent region of the IGM at this much higher redshift
\citep{z6.6}. Thus, the ionizing intensity during post-overlap must
have been inhomogeneous and must have increased gradually, with both
effects possibly caused by the shadows due to the remaining neutral
gas. Much of this gas may have been located inside the dense halos of
collapsed objects.

In the popular $\Lambda$CDM cosmology, most of the collapsed gaseous
halos prior to reionization had a virial temperature below $10^4$K, at
which cooling due to atomic transitions is heavily suppressed
[\citet{us01}, and references therein]. The gaseous low-mass halos
(hereafter ``minihalos'') thus remain intact until photoionization
heating by the cosmic UV background photo-evaporates much of this gas
back into the IGM during the overlap phase \citep{us99}. \citet{HAM}
noted that these halos may consume a significant number of ionizing
photons if they are assumed to be completely photo-ionized throughout
the time it takes them to expand back into the IGM. However,
\citet{C01} pointed out that as an ionizing front approaches a dense
gas halo, it slows down and is eventually halted outside the virial
radius, with a shock front entering the gas halo instead. Regardless
of the precise dynamics, the process of photoevaporation takes an
amount of time which is set by the characteristic sound speed of $\sim
10~{\rm km~s^{-1}}$ (corresponding to $T \sim 10^4$ K). Whether
minihalos play a substantial global role during this timespan depends
on their ability to screen ionizing sources.

The limited mass resolution of current numerical simulations of
reionization \citep{G00,R02} does not allow them to resolve these
minihalos and include their cumulative screening effect; \citet{HAM}
made a crude estimate of the covering factor of minihalos and found
that it could be significant. In this paper we gauge the importance of
absorption by minihalos with a detailed semi-analytic calculation. In
\S 2 we describe our model for the basic halo properties and the
statistics of their screening. In \S 3 we describe our numerical
results. Finally, \S 4 discusses the main conclusions of this
work. Throughout the paper we use the $\Lambda$CDM cosmological
parameters of $\Omm=0.3$, $\Omega_{\Lambda}=0.7$, and $\Omega_b=0.05$,
for the density parameters of matter, cosmological constant, and
baryons, respectively. We also assume a Hubble constant $h=0.7$ in
units of $100\mbox{ km s}^{-1}\mbox{Mpc}^{-1}$, and a primordial
scale-invariant ($n=1$) power spectrum with a value $\sigma_8=0.8$ for
the root-mean-square amplitude of mass fluctuations in spheres of
radius $8\ h^{-1}$ Mpc.

\section{Theoretical Model}

\subsection{Basic Halo Properties}

We begin with a description of our model for the properties of
minihalos and of ionizing sources in the context of hierarchical
galaxy formation in $\Lambda$CDM. The input for our opacity
calculation involves the distribution of halo masses for halos that
contain uncooled gas and for halos which host galaxies, as well as
knowledge of the density distribution inside each minihalo.

We assume that the abundance of halos is given by the model of
\citet{PS74}, with the modifications which \citet{st98} and
\citet{jenk00} used in order to fit more accurately the mass function seen 
in numerical simulations. Galaxies form in halos in which gas can
accumulate and cool. At high redshift, gas can cool efficiently in
halos down to a virial temperature of $\sim 10^4$ K or a circular
velocity of $V_c\sim 16.5\ {\rm km\ s}^{-1}$ with atomic cooling,
which we assume to be the dominant cooling mechanism (we consider the
effect of H$_2$ cooling in \S \ref{res}). Before reionization the IGM
is cold and neutral, and these cooling requirements set the minimum
mass for halos which can host galaxies. During reionization, the
ionized IGM is heated to a temperature $T_{\rm IGM}\sim 1$--$2 \times
10^4$ K, and gas infall into halos is suppressed by the increased
pressure force. However, this pressure suppression is not expected to
cause an immediate suppression of the cosmic star formation rate,
since even after fresh gas infall is halted the gas already in
galaxies continues to produce stars, and mergers among already-formed
gas disks also trigger star formation. We neglect the suppression
effect in this paper, since we consider only the era of reionization
and its immediate aftermath.

We use the term 'minihalos' to denote halos in which gas accretes but
is unable to cool. These halos contain virialized dark matter and gas,
where the gas does not collapse onto a disk and does not participate
in star formation. The maximum total mass of a minihalo is $1.6 \times
10^8 M_{\odot}$ at $z=6$, corresponding to the atomic cooling
threshold \citep[e.g.][]{us01}. This implies that the gas in minihalos
is almost entirely neutral prior to reionization. The minimum mass is
determined by the IGM temperature, which prevents gas from collecting
inside the smallest halos. The Jeans mass is the minimum mass-scale of
a perturbation for which gravity overcomes pressure, in linear
perturbation theory. The Jeans mass is related only to the evolution
of perturbations at a given time, while a collapsing halo is the end
result of the long evolution history of a perturbation. When the Jeans
mass varies during this time, the overall suppression of the growth of
perturbations depends on a time-averaged Jeans mass. \citet{gh98}
showed that the correct time-averaged mass is the filtering mass
$M_F$, which we adopt as the minimum mass of minihalos. At $z=6$, for
example, we find $M_F = 4.6 \times 10^4 M_{\odot}$ in areas that have
not yet been reionized and heated.

If gas collects inside a halo and also cools, it can collapse to high
densities and form stars (or possibly also a quasar). The ability of
stars to form is determined by gas accretion which, in a hierarchical
model of structure formation, is driven by mergers of dark matter
halos. In order to determine the lifetime of a typical source, we
first determine the time spent by gas (i.e. the gas `age') within a
given halo using the average rate of mergers which built up the
halo. Based on the extended Press-Schechter formalism \citep{lc93},
for a halo of mass $M$ at redshift $z$, the fraction of the halo mass
which by some higher redshift $z_2$ had already accumulated in halos
with galaxies is \be F_M(z,z_2) = {\rm erfc} \left(\frac{1.69/D(z_2)-
1.69/D(z)}{\sqrt{2 (S(M_{\rm min}(z_2))-S(M))}} \right)\ , \ee where
$D(z)$ is the linear growth factor at redshift $z$, $S(M)$ is the
variance on mass scale $M$ (defined using the linearly-extrapolated
power spectrum at $z=0$), and $M_{\rm min}(z_2)$ is the minimum halo
mass for hosting a galaxy at $z_2$ (determined by cooling as discussed
above).

We estimate the total age of gas in the halo as the time since
redshift $z_2$ where $F_M(z,z_2) = 0.2$, so that most ($80\%$) of the
gas in the halo has fallen into galaxies since then. Low-mass halos
form out of gas which has recently cooled for the first time, while
high-mass halos form out of gas which has already spent previous time
inside small galaxies. We emphasize that the age we have defined here
is not the formation age of the halo itself, but rather it is an
estimate for the total period during which the gas which is
incorporated in the halo has participated in star formation. However,
the rate of gas infall is not constant, and even within the galaxy
itself the gas likely does not form stars at a uniform rate. Indeed,
galaxies could go through repeated cycles of a star formation burst
followed by feedback squelching, followed by another cycle of cooling,
fragmentation and star formation. The details involve complex
astrophysics, which are not understood even at low redshift, so we
account for the general possibility of bursting sources by adding a
parameter $\zeta \le 1$, the duty cycle. When $\zeta < 1$, compared to
$\zeta=1$ there are fewer active sources at any given time (by a
factor of $\zeta$) but each has a larger star formation rate (by a
factor of $1/\zeta$). In addition, the star formation rate is
proportional to the efficiency parameter $\eta$, which is the fraction
of gas in galaxies which turns into stars.

Specifically, the ionizing source in a halo of mass $M$ emits, over
its entire lifetime (which at $z=6$ is typically $t_S \sim 1 \times
10^8 (\zeta/0.25)$ yr), the following number of photons per unit solid
angle: \be \frac{dP}{d\Omega} = 6.4 \times 10^{64}\, \left(
\frac{M}{10^8 M_{\odot}} \right)\, \left( \frac{N_{\rm ion}}{40}
\right)\, \left( \frac{\Omega_b/\Omega_m}{0.17} \right)\ . \ee In this
equation, $\Ni$ is the overall number of ionizing photons produced per
baryon in galactic disks; it includes 4000 ionizing photons produced
per baryon in stars [for a \citet{scalo} stellar IMF and a metallicity
equal to $1/20$ of the solar value\footnote{This low metallicity is
expected to roughly characterize $z\sim 6$ galaxies, which form out of
an IGM which may have been enriched by an even earlier generation of
galaxies; see, e.g., \citet{us01} for further discussion.}], times an
efficiency $\eta = 10\%$, times an escape fraction $f_{\rm esc}^{\rm
disk}=10\%$ of ionizing photons from the disk itself [see
\citet{us01} and references therein]. Our choice of typical
parameters $\zeta=0.25$ and $\eta=10\%$ is loosely based on numerical
simulations and on the low redshift cosmic star formation rate,
respectively, but the exact parameters valid during reionization can
only be determined from future observations.

\subsection{Screening due to Minihalos}
\label{screen}

At any given redshift we consider a single halo emitting ionizing
photons, surrounded by a distribution of mini-halos of various masses
at various distances $R$ from the center of the source halo. Except
where indicated otherwise, we neglect the ionizing radiation from
other sources when calculating the transmission of the radiation from
a source through the minihalos. This assumption is justified as long
as we consider the early stage of reionization when the ionizing
background has not yet been established, and as long as we consider a
small enough distance that the ionizing intensity is dominated by the
central ionizing source.

In order for the radiation to reach a large distance $R$, it must
overcome two obstacles caused by the minihalos: (i) it must ionize the
neutral gas within the minihalos, and (ii) it must overcome the
recombination rate of this gas in order to keep it ionized. Our
statistical treatment of both quantities is similar, so in general we
label the quantity we are considering by $\nu$. Thus, for example,
$\nu_1$ is the sum of the number of hydrogen atoms per unit solid
angle encountered by a line of sight going out to radius $R$. This sum
is then compared to the total photon emission per solid angle
$dP/d\Omega$ of the source, and the photons escape out to radius $R$
along those directions for which $dP/d\Omega > \nu_1$. Separately we
also calculate $\nu_2$, the sum of the number of recombinations per
second of hydrogen atoms (assuming the gas starts out ionized) per
unit solid angle along a line of sight out to radius $R$. Photons can
only escape along those directions for which $d^2 P/[d\Omega\, dt] >
\nu_2$, where the (average) emission rate of photons per solid angle
is \be \frac{d^2P}{d\Omega\, dt} = \frac{1}{t_S}\, \frac{dP}{d\Omega}\
, \ee in terms of the source lifetime $t_S$.

Both $\nu_1$ and $\nu_2$ are determined by the density distribution of
gas in the minihalos. Numerical simulations of hierarchical halo
formation indicate a roughly universal spherically-averaged density
profile for dark matter in the resulting halos (\citealt{NFW},
hereafter NFW), though with considerable scatter among different
halos. Such high-resolution simulations have not been done with gas.
However, in one-dimensional simulations of dark matter and adiabatic
gas, we find [based on the methods of \citet{TW} and \citet{HTL96}]
similar final density profiles for the gas and dark matter, where both
components form a cusp and the gas profile is only slightly shallower
than the dark matter profile. Note that this result is found even
though the physics of the two components is rather different, with the
dark matter undergoing violent relaxation due to shell crossing, while
the gas is halted at a virialization shock which moves outward. For
simplicity, we therefore assume that even in the three-dimensional
case the virialized gas profile is similar to that of the dark matter,
and so we adopt the radial NFW profile for the gas density in
minihalos, \be \rho(r)=\frac{\rho_0} {\cN x (1+\cN x)^2}\ ,
\label{NFW} \ee where $x=r/r_{\rm vir}$, $\cN$ is the concentration 
parameter, and the virial radius is \be r_{\rm vir} = 2.1
\left(\frac{M}{10^8 M_{\sun} }\right)^{\frac{1}{3}}\,
\left(\frac{\Omm} {0.3}\right)^{-\frac{1}{3}}\,
\left (\frac{1+z}{7} \right)^{-1}\, \left( 
\frac{h}{0.7} \right)^{-\frac{2}{3}}\, {\rm kpc}\ ,\ee where
in this equation and in what follows we assume a sufficiently high
redshift for the Einstein-de Sitter limit [$\Omega_m(z) \approx
1$]. The value of the concentration parameter for minihalos at high
redshift is unknown. We extrapolate from simulations of \citet{cN},
which indicate
\be \cN \approx 4.5\, \left(\frac{M}{10^8 M_{\sun}}\right) ^{-0.1}\, 
\left(\frac{1+z}{7}\right)^{-1} \ . \ee For the population of minihalos 
we adopt for simplicity a single value of $\cN$, the value expected
for a minihalo of mass equal to the mass-weighted mean minihalo mass.

Consider a line of sight which passes through a minihalo of mass $M$,
with an impact parameter of $\alpha$ (in units of $r_{\rm vir}$)
relative to the center of the minihalo. The total hydrogen column
density is then \be N_{\rm H I} = 5.6 \times 10^{19}\, \frac{\cN^2}
{g(\cN)}\, f_1(\alpha,\cN,x_m)\, \left(
\frac{M}{10^8 M_{\odot}} \right)^{1 \over 3}\, \left(
\frac{1+z}{7} \right)^2\, \left(\frac{\Omega_b}{0.05}\right)\, 
\left( \frac{\Omega_m}{0.3} \right)^{-\frac{1}{3}}\, \left( 
\frac{h}{0.7} \right)^{\frac{4}{3}}\,{\rm cm}^{-2}\ , \label{eq:NHI}
\ee 
where we include gas out to a maximum radius $x_m$ (in units of
$r_{\rm vir}$), and where \be g(\cN) = \log(1+\cN)-\frac{\cN}{1+\cN}\
.\ee Defining \be \omega_1=1-\alpha^2 \cN^2 \ee and \be \omega_2=1+\cN
x_m \ , \ee we derive the function
\be
f_1(\alpha,\cN,x_m) = -\frac{\cN}{\omega_2}\, \frac{ \sqrt{
x_m^2-\alpha^2}} {\omega_1}\, + \, \frac{1}{|\omega_1|^{3 \over2}}\,
\times\, \left\{ \begin{array}{ll}
\log \left| \frac{\alpha \omega_2} {\alpha^2 \cN + x_m -
\sqrt{\omega_1(x_m^2-\alpha^2)}} \right| & {\rm if}\  
\alpha < {1 \over \cN} \\*[0.15 in] \sin^{-1}\left[\frac{\alpha^2 
\cN + x_m}{\alpha \omega_2} \right] - {\pi \over 2} & 
\mbox{otherwise.} \end{array} \right.
\ee

Similarly, the total recombination rate per unit area, along the
same line of sight, is 
\be 
\Gamma_{\rm rec} = 5.4 \times 10^{4}\, \frac{\cN^4} {g^2(\cN)}\, 
f_2(\alpha,\cN,x_m)\, \left(
\frac{M}{10^8 M_{\odot}} \right)^{1 \over 3}\, \left(
\frac{1+z}{7} \right)^5\, \left( \frac{\Omega_b}{0.05}
\right)^2\, \left( \frac{\Omega_m}{0.3} \right)^{-\frac{1}{3}}\,
\left( \frac{h}{0.7} \right)^{\frac{10}{3}}\, 
{\rm cm}^{-2}\, {\rm s}^{-1}\ . \ee 
Using $\omega_1$ and $\omega_2$ from above and defining \be
\omega_3=1- \frac{2}{\omega_2} + \frac{\omega_1}{\omega_2^2}\ ,\ee
we find
\ba
f_2(\alpha,\cN,x_m) & = & 
\frac{\sqrt{\omega_3}}{\omega_1}\, \left(1+\frac{1}{2 \omega_2}+
\frac{3}{2 \omega_1} \right)+ \frac{{\pi \over 2}-\sin^{-1}
\left({\alpha \over x_m}\right)}
{\alpha \cN} + \left(1+\frac{1}{2 \omega_1}+\frac{3}{2 \omega_1^2}
\right) |\omega_1|^{-\frac{1}{2}}
\, \nonumber \\
& & \times \left\{ \begin{array}{ll}
\log \left| \frac{1-\frac{\omega_1}{\omega_2}- \sqrt{\omega_1 
\omega_3}}{\alpha \cN}\right| & {\rm if}\  
\alpha < {1 \over \cN} \\*[0.15 in] \sin^{-1}\left[
\frac{\alpha \cN + x_m/\alpha}{\omega_2} \right] - 
{\pi \over 2} & \mbox{otherwise.} \end{array} \right.
\ea

\subsection{Statistics of Screening}
\label{stats}

The quantities $\nu_1$ and $\nu_2$ introduced in the previous
subsection are sums over the contributions from all minihalos
encountered along a line of sight, where a single cross-section of a
minihalo at proper distance $R$ contributes $N_{\rm H I} R^2$ to
$\nu_1$ and $\Gamma_{\rm rec} R^2$ to $\nu_2$. We wish to calculate
the optical depth $d \tau$ for getting a value in the range $\nu_1$ to
$\nu_1+d\nu_1$ by intersecting a minihalo along a line of sight. This
involves an integral along the line of sight of the minihalo abundance
times cross-sectional area. We first note that the cross-sectional
area of a single minihalo, for a given fractional impact parameter
$\alpha$, is $\pi \alpha^2\, r_{\rm vir}^2$. Thus, the contribution to
a given value of $\nu_1$ is $2 \pi
\alpha\, d
\alpha\, r_{\rm vir}^2$; determining the range $d \alpha$ that
contributes to a given range $d \nu_1$, we arrive at the total optical
depth:
\be 
\nu_1 \frac{d \tau}{d\nu_1} = \int_0^{R_{\rm max}} dR
\int_{M_{\rm min}}^{M_{\rm max}} dn(M,z)\ b(R,M,z)\, 2 \pi 
\alpha r_{\rm vir}^2(M,z)\, \frac{f_1(\alpha,\cN,x_m)}{\left| 
\frac{\partial} {\partial \alpha} f_1(\alpha,\cN,x_m)\right|}\ ,
\ee
where $dn$ is the halo abundance, $b(R,M,z)$ is the bias of minihalos,
and everywhere in the integrand $\alpha$ is determined by inverting
the equation $N_{\rm H I}(M,z,\alpha) R^2 = \nu_1$. We get a similar
formula for $\nu_2$ as well.

A crucial component of the minihalo abundance is the bias $b(R,M,z)$,
which accounts for the biased distribution of minihalos near the
source halo. We discuss here the calculation of bias in some
detail. First, we set the bias equal to zero within the virial radius
of the source halo, since minihalos cannot exist inside the source
halo. Outside the virial radius, the bias can be divided into two
components. One can be called ``Lagrangian bias'', or ``initial
condition bias'', and results from the fact that the initial
perturbations in nearby points are correlated. Although various
analytical approaches to Lagrangian correlations have been suggested,
we use the natural and general approach of \citet{mePast}, who worked
directly within the standard theoretical framework of excursion sets
\citep{B91,lc93}. Their results yield the bias of two halos at a given
distance as a function of the two halo masses and the two formation
redshifts (although in this application we require only the equal
redshift case). Most importantly for the current application, this
approach to bias explicitly accounts for the existence of the source
halo. For example, there is a strong correlation between nearby rare
peaks in the initial density field, which tends to give a positive
bias among massive halos at high redshift; on the other hand, all
halos are anti-biased at initial positions that are very close to a
given halo $A$, since much of the mass there is likely to belong to
the halo $A$ itself when it finally forms. Because of this effect,
minihalos are anti-biased at initial positions that are close to a
source halo, while source halos themselves are positively correlated
at the same distances (though both are anti-biased at the very
smallest distances).

The second part of the bias can be called ``Eulerian bias'' or
``infall bias'', and is a result of the high-density environment
around the massive source halo when it forms, due to infall of mass
from a large initial region in the direction of the forming halo. If,
for example, minihalos contain 25\% of the mass density in an unbiased
region, then in the absence of Lagrangian bias, minihalos would
contain 25\% of the mass density also near the source halo; thus, the
absolute number density of minihalos would increase in proportion to
the general mass overdensity near the source halo. In general, the
Lagrangian bias must be included as an additional factor beyond
this. In the spherical approximation, the Eulerian bias depends only
on the final density profile near the virialized source halo; assuming
there is no shell crossing (a reasonable approximation outside the
virial radius), we can map each mass shell in the final density
profile to its initial ``Lagrangian'' position, and we then use this
initial position in order to calculate the Lagrangian bias
factor. There are a number of possible approaches to estimate the
density profile outside the virial radius of the source halo, but
approaches based on spherical collapse, for example, do not match the
profile from numerical simulations near the virial radius. However,
all approaches give three-dimensional density profiles with power-law
slope between $R^{-2}$ and $R^{-3}$, and in any case, realistically we
expect a large scatter in the profiles around halos. For simplicity,
here we extrapolate the NFW profile of the source halo to larger
radii, an approach which also gives a fall-off between $R^{-2}$ and
$R^{-3}$ in the relative overdensity (i.e., we assume that at large
$R$ the density approaches its cosmic mean value).

In order to calculate the fraction of ionizing photons from the source
that escape out to radius $R$, we require the probability
distributions of $\nu_1$ and $\nu_2$. However, the above optical depth
$d\tau$ does not directly yield the probability; it only accounts for
the possibility of intersecting a single minihalo, while there can be
multiple intersections along a single line of sight. In addition, the
total optical depth can be greater than unity. We determine the actual
probability distribution with a Monte Carlo approach. Specifically, we
divide the total optical depth to $\sim 100$ bins, each of which
contains a $\Delta \tau \ll 1$. In each Monte Carlo trial, we obtain a
total $\nu_1$ (for example) by summing over the bins; a given bin
makes a contribution with probability $\Delta \tau$, and the
contributed value is determined by the distribution $d\tau/d\nu_1$
within the bin. We determine the probability distribution using $\sim
100,000$ trials (for each set of parameters). Note that our
calculation assumes that the biasing of the minihalo abundance is
determined by the correlation with the nearby source halo, and we
neglect any additional correlations among the minihalos themselves.

Once we have obtained the probability distributions of $\nu_1$ and
$\nu_2$, we can calculate the probability distribution of the escape
fraction of photons. We often find it convenient to summarize this
distribution with a single quantity, the mean escape fraction of
photons, averaged over the probability distribution (which corresponds
to averaging over solid angle). Including only the constraint of
initial ionization (i.e., $\nu_1$), the escape fraction is
\be \label{eq:fesc}
f_{\rm esc}^{(1)} = \int_{\nu_1=0}^{dP/d\Omega} 
\left( 1- \frac{\nu_1} {dP/d\Omega} \right)\, \frac{dp}{d\nu_1}\, 
d\nu_1\ , \ee where $dp/d\nu_1$ denotes the probability distribution
of $\nu_1$.  We similarly get $f_{\rm esc}^{(2)}$, with $\nu_1$ and
$dP/d\Omega$ replaced by $\nu_2$ and $d^2P/[d\Omega\, dt]$,
respectively. Combining the two constraints exactly would be difficult
given the fact that they are correlated, since a given minihalo
contributes to both $\nu_1$ and $\nu_2$; in practice, we estimate the
final escape fraction by taking the lower of the two, which is usually
$f_{\rm esc}^{(2)}$.

\subsection{Characteristic Distances}
\label{whichR}

When we consider the escape of photons out to various distances from a
source, some values of distance are particularly important. We first
want to determine a fiducial distance $R$ that represents the
beginning of the overlap stage of reionization. The overlap stage
starts once neighboring \ion{H}{2} regions begin to merge, thus
increasing the ionizing intensity within these regions because of the
access to multiple ionizing sources. The rising intensity allows the
\ion{H}{2} regions to expand faster and into denser regions, leading
to additional mergers, with the process accelerating until the entire
IGM becomes highly ionized except for gas located inside
self-shielded, high-density clouds.

One guess for the distance $R$ which characterizes the beginning of
overlap is simply the typical distance to the nearest other ionizing
source. However, this value of $R$ does not properly represent the
beginning of overlap; first, it may be too large since the photons
need not reach the nearby source itself but only the \ion{H}{2} region
produced by that source; and second, this $R$ may be too small, since
if the nearby source has a mass much smaller than that of the first
source then it will make a negligible addition to the ionizing
intensity and this will not advance the process of overlap. We thus
choose a distance $R$ which more accurately reflects these
considerations. First we find that distance where we would typically
expect the total mass in additional ionizing sources to be equal to
the mass $M$ of the first source\footnote{We avoid the case where the
average mass $M$ corresponds to a halo of mass $M_2 \gg M$, which is
present only rarely (with probability $M/M_2$); thus, we choose the
distance where we expect to either find a single halo with $M_2 > M$,
or a total mass $M$ in source halos for which $M_2 < M$
individually.}. Note that we only count active sources (when the duty
cycle $\zeta < 1$). Now, since the other sources also send photons
toward the first source, we take only a fraction of the distance
(where just the portion beyond the virial radius is reduced). The
precise fraction depends on the actual masses of the additional
sources; we adopt a fraction of $1/2$, which is exact for the
symmetric case of having a second source halo of mass equal to
$M$. This final choice of proper distance $R$, which represents the
beginning of overlap, is referred to hereafter as half the equal-mass
distance.

A second important distance is related to the Gunn-Peterson trough.
The essential feature that defines Gunn-Peterson absorption is the
presence of long stretches of continuous, strong \Lya absorption. Such
absorption has been revealed in recent observations at $z \sim 6$
\citep{z6.3,z5.8}. Neutral hydrogen absorbs photons near the
\Lya resonance with a very high opacity as given by \citet{GP}.
As a result, the IGM itself produces strong absorption unless it is
highly ionized. The IGM is highly ionized only if the ionizing
intensity is sufficiently high. For a given distribution of ionizing
sources, each point in the IGM must see sources out to some distance
$R$ in order to reach the necessary total ionizing intensity. This
proper distance $R$ is given by [see equation (18) of \citet{z6me}]
\be R_{\rm GP} = 3.4
\left(\frac{\tau_{{\rm Ly}\, \alpha}} {2.5}\right)^{-1}\,
\left({1+z\over 7} \right)^{3 \over 2}\,
\left({\Omega_m\over 0.3} \right)^{-{1 \over 2}}\,
\left({\Omega_b h\over 0.035} \right)\,
\left( \frac{t_S}{5 \times 10^8\mbox{ yr}} \right)\, 
\left( \frac{\Ni}{40}\, \frac{F_{\rm col}} {0.1}\right)^{-1}\,
\left( \frac{\Delta_{\tau}}{0.4}\right)^2\, {\rm Mpc}\ .
\label{eq:z6me}
\ee
We have defined here the end of the Gunn-Peterson trough as the time
when the optical depth drops below $\tau_{{\rm Ly}\, \alpha}=2.5$. The
total gas fraction inside source halos is $F_{\rm col}$. For the
purposes of this estimate we have also used the approximation that all
sources have the same lifetime $t_S$, where everywhere in this
equation we have assumed $\zeta=1$ since the value of $\zeta$ does not
affect the total ionizing intensity. The density in voids which
typically dominates the \Lya transmission is $\Delta_{\tau}=0.4$ in
units of the cosmic mean gas density. We have neglected source
clustering, but such clustering only makes voids relatively empty of
sources, increasing the distance $R$ required to prevent
transmission. For additional discussion of equation (\ref{eq:z6me}),
see \citet{z6me}.

When we investigate the end of the Gunn-Peterson trough, we must also
consider the effect of the ionizing background in partially ionizing
the minihalos. This is required because the end of the Gunn-Peterson
trough occurs relatively late, after an ionizing background may have
already been established; furthermore, it is difficult for the central
ionizing source to dominate the ionizing intensity at the rather large
distance $R_{\rm GP}$. We estimate the effect of the ionizing
background as follows. First, we find the distance $R_{0.5}$ at which
$f_{\rm esc}=0.5$ for a typical source (i.e., a source with mass equal
to the mass-weighted mean mass of all sources). We assume that the
ionizing background at a random point in the IGM is given by the
average integrated intensity due to all sources out to the distance
$R_{0.5}$, times one half (since $f_{\rm esc}=0.5$ at this distance).
With this background intensity, ionization equilibrium yields the
neutral fraction at each radius of a given minihalo. For simplicity,
we find the neutral fraction at the virial radius of a typical
minihalo, and we then assume this neutral fraction to apply throughout
every minihalo. This overestimates the effect of the ionizing
background, since the central portions of minihalos are in reality
denser and more highly neutral; however, as shown in \S \ref{res},
even with this overestimate the overall effect of the partial
ionization on screening is small. Once partial ionization is included,
$f_{\rm esc}$ increases at a given $R$, and thus $R_{0.5}$ also
increases; we iterate in order to find the self-consistent value of
$R_{0.5}$, and the corresponding value of the neutral fraction in
minihalos. We then use this neutral fraction when we compute screening
at $R=R_{\rm GP}$.

\subsection{Direct Gunn-Peterson Absorption}
\label{Lya}

In addition to their screening of ionizing photons, minihalos also
absorb \Lya photons and may therefore contribute to the Gunn-Peterson
trough. Direct absorption near the \Lya resonance is particularly
strong (\citealt{GP}; see also the related discussion in the previous
subsection). However, minihalos contain $\sim 20\%$ of the gas in the
universe at an average overdensity of $\sim 200$; since for any
population of absorbers the covered fraction of a line of sight equals
the volume fraction of the population\footnote{More precisely, the
covering factor of a line of sight equals the volume filling
factor. Consider, for simplicity, a number density $n$ of halos which
all have the same shape and inclination to the line of sight. A single
halo has a cross-sectional area $A=\int_A dx\, dy$, where we use
two-dimensional $(x,y)$ coordinates. Denote the total depth of the
halo at each point $D(x,y)$. On a random line of sight, the halos
cover a fraction $= n \int_A dx\, dy\, D(x,y)$. This integral equals
the volume of a single halo. This proof is easily generalized to a
population of halos with various shapes.}, minihalos directly cover
only $\sim 0.1\%$ of a random line of sight to a quasar. More
interesting, therefore, is the possibility of absorption due to the
damping wings of the \Lya line \citep{jordi98}.

To estimate the probability that the population of minihalos produces
strong absorption, we first consider the optical depth due to a single
minihalo; we do this as a function of distance $R$ from the minihalo,
where the distance represents different wavelengths $\lambda$ by the
correspondence of $\lambda$ and $z$ via a pure Hubble velocity flow
[see the similar approach, e.g., in \citet{LR99}]. We first note a
number of simplifications: the size of each halo (typically a few kpc)
is much smaller than the region of significant absorption due to the
halo's damping wings (typically a fraction of a Mpc); the latter
region, in turn, is much smaller than the horizon at the reionization
redshift. Under these conditions, the damping wing produces an optical
depth $\tau_{{\rm Ly}\, \alpha}$ which is simply proportional to the
total column density of \ion{H}{1} divided by the distance squared
from the minihalo. This results from the Lorentzian damping profile of
the line, and it corresponds to the well-known result that when the
damping wings dominate, the equivalent width of an absorption line is
proportional to the square root of the column density (note that at
small distances from the absorber, distance is proportional to change
of wavelength $\Delta
\lambda$). Specifically, at a distance $R$ from a column density
$N_{\rm H I}$,
\be
\tau_{{\rm Ly}\, \alpha} = 0.51 \left( \frac{N_{\rm H I}}{10^{20}\, 
{\rm cm}^{-2}} \right) \left( \frac{R} {1\, {\rm Mpc}} \right)^{-2}
\left( \frac{1+z} {7} \right)^{-3} \left( \frac{\Omm} {0.3} \right)^{-1} 
\left( \frac{h} {0.7} \right)^{-2}\ .
\label{eq:Lya}
\ee

Including the damping wings of the entire population of minihalos, we
calculate the total optical depth (or covering factor) $C$ along the
line of sight, for getting a \Lya optical depth above some value
$\tau_{{\rm Ly}\, \alpha}$. The cross-sectional area contributed by
the impact parameter range $d\alpha$ is $2 \pi \alpha\, d \alpha\,
r_{\rm vir}^2$; for a given $\alpha$, if we determine the maximum
distance $R$ from the minihalo for which the optical depth is greater
than $\tau_{{\rm Ly}\, \alpha}$, then the minihalo covers with this
optical depth a total line-of-sight distance $2R$ (including the blue
and red damping wings). Thus, the covering factor equals
\be
C(> \tau_{{\rm Ly}\, \alpha}) = \int_0^{x_{\rm max}} 
d\alpha \int_{M_{\rm min}}^{M_{\rm max}} dn(M,z)\ 4 \pi
\alpha r_{\rm vir}^2(M,z)\, R\ ,
\label{eq:covLya}
\ee
where $R$ in the integrand is determined from $\tau_{{\rm Ly}\,
\alpha}$ by inverting equation (\ref{eq:Lya}) for each $M$ and
$\alpha$, using equation (\ref{eq:NHI}) for $N_{\rm H I}$. As noted in
\S \ref{screen}, we also adopt a single concentration parameter $\cN$
for all minihalos; this gives them similar density profiles, which
simplifies the calculation (i.e., the double integral in equation
(\ref{eq:covLya}) can then be factored into two single-variable
integrals). Note that the covering factor is inversely proportional to
$\sqrt{\tau_{{\rm Ly}\, \alpha}}\,$. 

As in \S \ref{stats}, here too the optical depth includes only the
contribution from one minihalo at a time, and we determine the full
probability distribution of values of $\tau_{{\rm Ly}\, \alpha}$ with
Monte Carlo trials. Note that in this calculation, we do not include
clustering among the minihalos themselves. Clustering will change the
probability distribution somewhat, increasing the optical depth in
regions of clustered halos while decreasing the optical depth in
relatively empty regions. However, the damping wings effectively
smooth the absorption on the scale of a few comoving Mpc, reducing the
effect of clustering. There is a similar smoothing of the effect of
peculiar velocities, which we likewise neglect.

\subsection{Photoevaporation of Minihalos}
\label{evap}

As discussed above, the minihalos can affect the evolution of the IGM
in a number of ways; the gas, however, does not remain in minihalos
forever. \citet{us99} showed that photoionization heating by the
cosmic UV background can evaporate much of this gas back into the IGM
during the overlap era of reionization. This work also showed that
this process affects a broad range of halo masses, with only a small
gas fraction evaporating out of $10^8 M_{\sun}$ halos, but with halos
below $\sim 10^6 M_{\sun}$ losing their entire gas content because of
their shallow gravitational potential wells. Even after the initial
loss of unbound gas, the bound gas in the minihalo expands and may
further evaporate; however, this stage proceeds more slowly.

In this paper we consider two main stages in this process. In the
first stage, before reionization has begun, the minihalos are fully
intact. In the second stage, only the bound gas remains, a situation
which we approximate by keeping only the minihalos in which $> 50\%$
of the gas is initially bound after UV heating; this gives a minimum
circular velocity $V_c \sim 11~{\rm km~s^{-1}}$ which is independent
of redshift \citep{us99}. In the transition from the first to the
second stage, the typical mass of the minihalos that photoevaporate
(i.e., their mass-weighted mean) is $7 \times 10^6 M_{\sun}$. Such a
halo has a virial radius of 0.9 kpc, and once it is heated it should
evaporate with a $10~{\rm km~s^{-1}}$ outflow in $\sim 90$ million
years, which is equivalent to the time between $z=6.5$ and
$z=6$. Additional evolution during the second stage is even slower;
the expansion of the remaining minihalos (typical $M \sim 9 \times
10^7 M_{\sun}$) takes place on a timescale of $\sim 200$ million years
(equivalent to the redshift interval between $z=7.2$ and $z=6$), and
during this time each minihalo is expected to accrete additional dark
matter or merge with other minihalos. In any case, it is clear that if
minihalos obstruct reionization then their relatively slow
photoevaporation can stretch the reionization era over a significant
redshift range, as we discuss further below.

The two-stage process that we have defined is only an approximation
and it neglects some complications. For example, when an ionizing
source begins to ionize the minihalos around it, those minihalos can
immediately begin to photo-evaporate; however, as just discussed, most
of the gas in minihalos takes a substantial amount of time to
photoevaporate, even in the first stage which involves the unbound
gas. Our calculations apply during this period.

\section{Results}
\label{res}

\begin{figure}[htbp]
\plotone{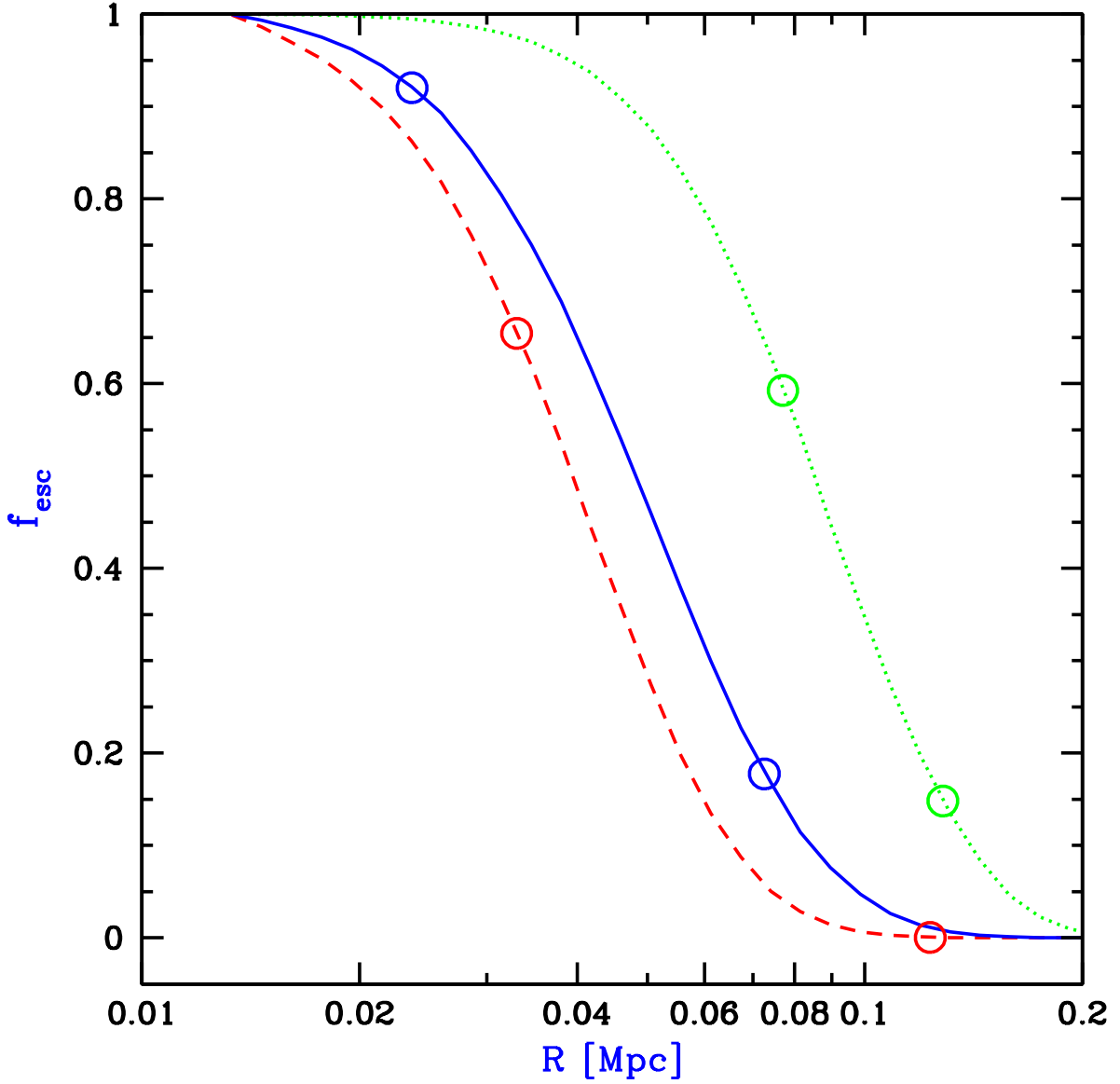}
\caption{$f_{\rm esc}$ versus physical distance $R$ from source halo. 
All curves assume $z=6$ and the mass-weighted mean source halo mass
$M=2.2 \times 10^{10} M_{\odot}$. Shown is $f_{\rm esc}$ assuming no
bias (dotted curve), infall bias (dashed curve), and full bias (solid
curve). On each curve, two distances are marked with open circles;
these are the distance to the nearest ionizing source (left) and half
the equal-mass distance (right; see \S \ref{whichR}).}
\label{fig-fOfR}
\end{figure}

Our main goal is to determine quantitatively the degree of effective
screening of ionizing sources due to the minihalo population. Figure
\ref{fig-fOfR} shows the escape fraction $f_{\rm esc}$ of ionizing
photons as a function of the physical distance $R$ from the source. We
pick a representative of the population of ionizing sources, namely a
halo of mass $M=2.2 \times 10^{10} M_{\odot}$, which is the
mass-weighted mean source halo mass (assuming that the minimum halo
mass for sources is $1.6 \times 10^8 M_{\odot}$ as determined by
atomic cooling). As shown in the figure, the typical distance from
this halo to the nearest other ionizing source is $R = 24$ kpc, giving
a high $f_{\rm esc} = 92\%$. However, as discussed in \S \ref{whichR},
the distance which best represents the beginning of overlap is half
the equal-mass distance. For the case shown in the figure, this
distance is $R = 73$ kpc, giving a low $f_{\rm esc} = 18\%$.

Figure \ref{fig-fOfR} also illustrates the importance of a full
calculation of the biased distribution of minihalos and of other
source halos around the first source halo. If we did not include bias,
we would calculate half the equal-mass distance to be $R=0.13$ Mpc,
and $f_{\rm esc}=15\%$. Including only the infall bias (see \S
\ref{stats}) increases the abundance of both minihalos and source 
halos\footnote{When the second halo mass $M_2 > M$, infall bias is
computed using infall onto $M_2$. Also, regardless of the choice of
bias, $M$ is not allowed to fall within the virial radius of $M_2$.}.
However, $R$ is reduced only slightly to 0.12 Mpc, since at this large
distance from the virial radius, the total enclosed mass is
essentially unchanged by infall; however, $f_{\rm esc} = 0.03\%$ is
still strongly reduced at these large distances because of the
screening due to the large concentration of minihalos very close to
the source. To better understand this, consider moving a given
minihalo twice closer to the source; the resulting halo then
contributes 4 times the covering factor, but produces values of
$\nu_1$ and $\nu_2$ that are reduced by $1/4$. We indeed find that
infall bias increases the covering factor most significantly at the
lowest values of $\nu_1$ and $\nu_2$. Finally, when Lagrangian bias is
included, $R$ is further reduced due to the strong positive
correlation among source halos; $f_{\rm esc}$, however, actually
increases (at a given $R$) because of a slight anti-bias of minihalos
around the source halo (see \S \ref{stats}).

\begin{figure}[htbp]
\plotone{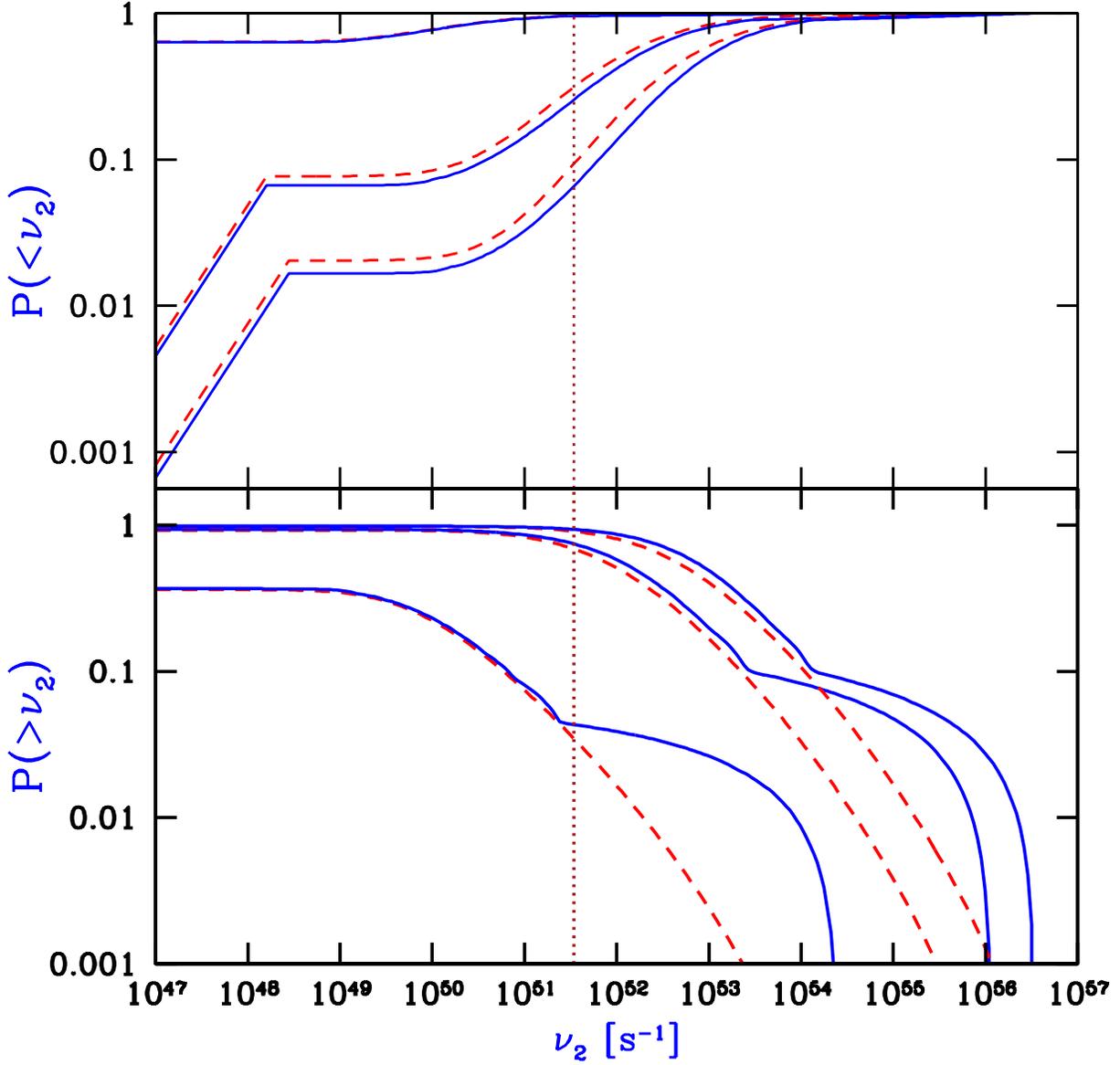}
\caption{Cumulative probability $P(>\nu_2)$ (lower panel) and 
corresponding $P(<\nu_2)=[1-P(>\nu_2)]$ (upper panel), versus total
recombination rate along the line of sight $(\nu_2)$. Shown are the
naive probability derived from the covering factor (dashed curves),
and the actual probability derived with a Monte Carlo approach (solid
curves). Also indicated is the rate of production of ionizing photons
by the source (vertical dotted line). The parameters correspond to the
solid curve in Figure \ref{fig-fOfR}, at three different radii. In
increasing order, the first two radii are the left and right points
indicated on the curve in Figure \ref{fig-fOfR}, and the third radius
is $R=0.1$ Mpc. (Increasing $R$ corresponds to increasing $P(>\nu_2)$
and decreasing $P(<\nu_2)$.)}
\label{fig-POfnu}
\end{figure}

Figure \ref{fig-POfnu} shows the full probability distribution of
$\nu_2$, from which $f_{\rm esc}$ (shown in Figure \ref{fig-fOfR}) is
derived using equation (\ref{eq:fesc}). The figure compares the
results of the Monte Carlo calculation, which includes the effect of
screening due to multiple minihalos, to the naive probability $P(>
\nu_2) = 1-e^{-\tau}$, given an optical depth $\tau$ for
having a recombination rate per unit solid angle greater than $\nu_2$.
Clearly, the effect of multiple screening changes completely the
probability distribution at high values of $\nu_2$; however, whenever
the probability of screening the source is significant, it is
dominated by the contribution of relatively low values of $\nu_2$, and
the mean $f_{\rm esc}$ is only moderately affected by multiple
scattering. For example, for the three cases shown in the figure, the
naive probability gives $f_{\rm esc}=5.8\%$, $21\%$, and $93\%$ (in
order of increasing $R$), while the exact probability distribution
yields $f_{\rm esc}=4.2\%$, $18\%$, and $92\%$, respectively. Note
also that our inclusion of partial blocking as in equation
(\ref{eq:fesc}) is crucial for getting the correct value of $f_{\rm
esc}$. For example, using the total covering factor of all values of
$\nu_2$, the naive probability of having no blocking at all is
$2.0\%$, $7.7\%$, and $64\%$, respectively; this probability is not an
accurate estimate for $f_{\rm esc}$.

\begin{figure}[htbp]
\plotone{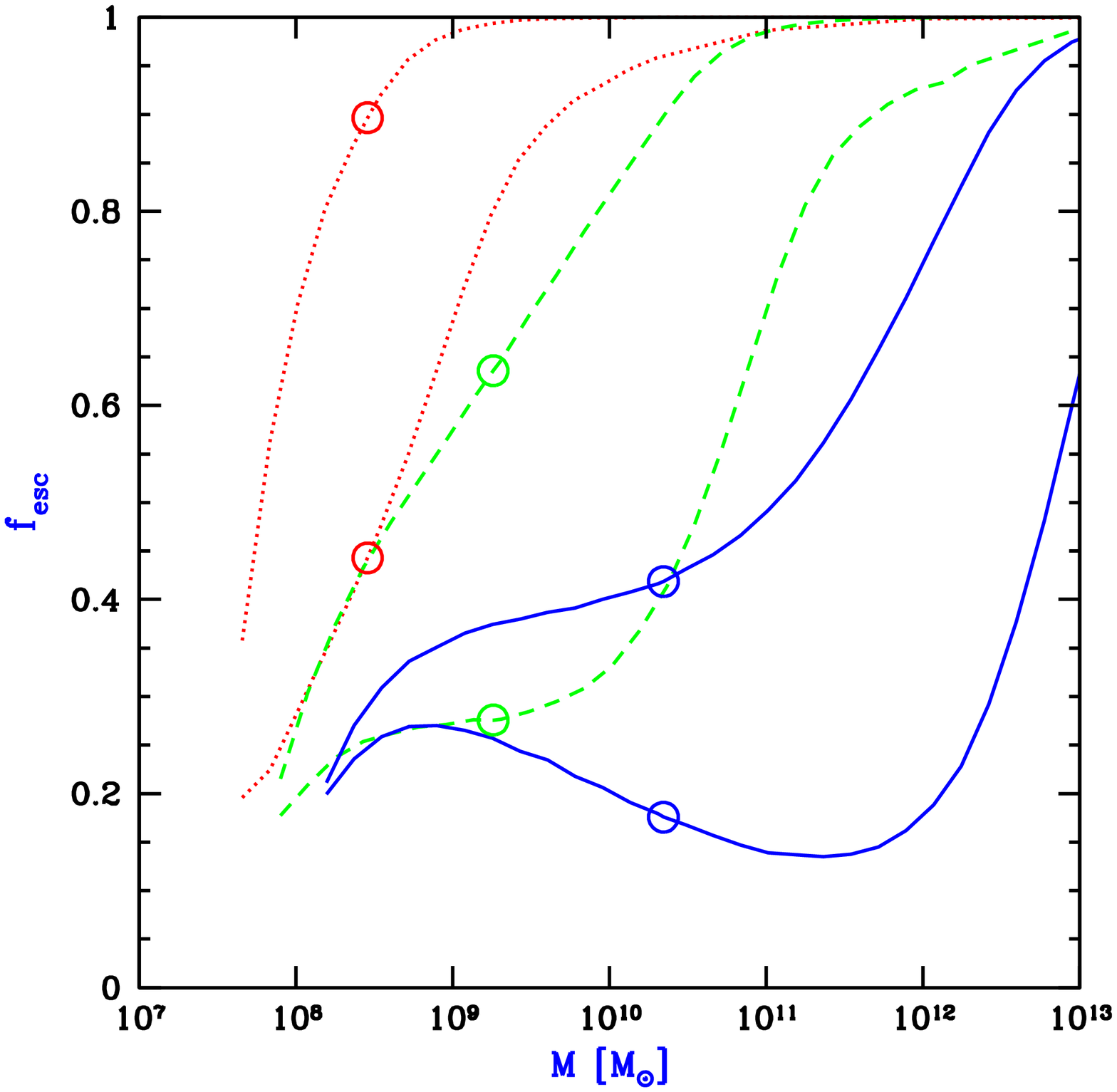} 
\caption{$f_{\rm esc}$ versus source halo mass $M$. Shown is $f_{\rm esc}$ 
at half the equal-mass distance (see \S \ref{whichR}), for the
populations of sources and minihalos at $z=6$ (solid curves), $z=10$
(dashed curves), and $z=15$ (dotted curves). In each pair, the upper
curve corresponds to $\nu_1$ (constraint of initial ionization), and
the lower curve corresponds to $\nu_2$ (constraint of
recombinations). An open circle on each curve marks the mass-weighted
mean source halo mass, where the minimum source halo mass is assumed
to be determined by atomic cooling. All curves assume our case of full
bias.} \label{fig-fOfM}
\end{figure}

Although Figure \ref{fig-fOfR} illustrates the results for a single
representative source halo mass, a wide range of halo masses may in
fact contribute to reionization. Figure \ref{fig-fOfM} illustrates how
the effectiveness of screening varies with halo mass.  The variation
with halo mass is complicated by several competing effects. On the one
hand, larger halos produce a higher flux of ionizing photons. On the
other hand, larger halos must in general send their photons farther
before they can effectively participate in overlap, i.e., there is an
increase with halo mass in the equal-mass distance; however, the rate
of increase declines at the high-mass end because of the powerful
clustering of extremely massive halos. The end result is that for the
case of $z=6$, minihalos screen all source halos up to $2 \times
10^{12} M_{\odot}$ roughly equally, with $13\% < f_{\rm esc} < 27\% $;
although halos with even higher masses are not strongly screened, most
ionizing photons are produced by halos for which $f_{\rm esc}$ is low.

Figure \ref{fig-fOfM} also shows the separate escape fractions which
result from the obstacles (see \S \ref{screen}) caused by initial
ionizations ($\nu_1$) and by recombinations ($\nu_2$). In each case,
recombinations give the stronger constraint, hence we have generally
used recombinations to determine escape fractions. The figure shows as
well the dependence of $f_{\rm esc}$ on redshift, where we make the
individual sources more efficient at high redshift, so that the total
production rate of ionizing photons remains fixed, and thus
reionization can occur at the higher redshift. At high redshift, the
total mass fraction in minihalos is smaller, but the biasing around
the source is stronger both for minihalos and for other sources. The
final result is that for a typical source (whose mass declines with
redshift), $f_{\rm esc}$ increases with redshift, from $18\%$ at $z=6$
to $28\%$ at $z=10$ and $44\%$ at $z=15$.

\begin{figure}[htbp]
\plotone{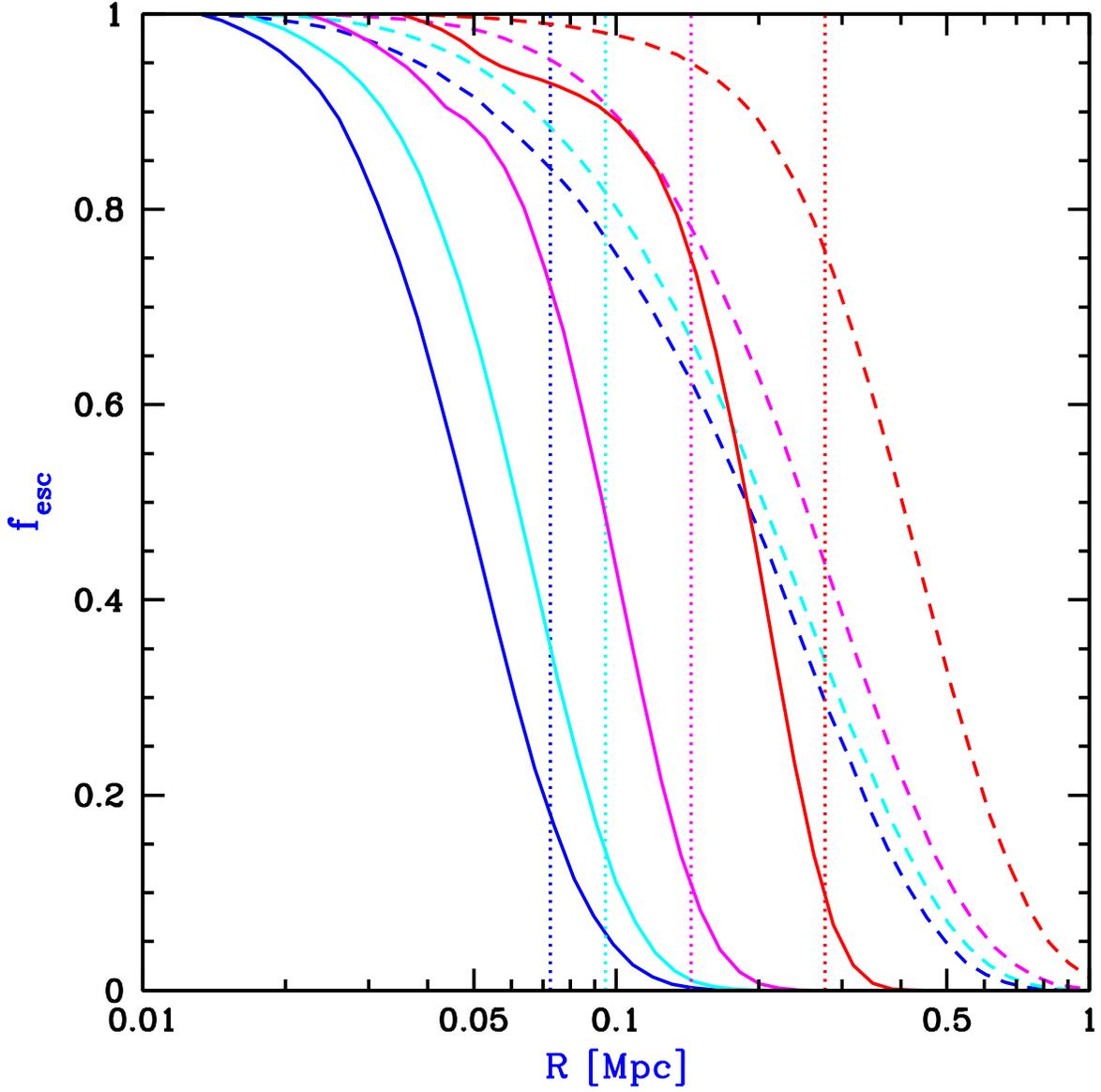}
\caption{$f_{\rm esc}$ versus physical distance $R$ from source halo. 
Shown are $f_{\rm esc}$ for all mini-halos (solid curves), and for
mini-halos in which most of the gas remains bound (dashed curves; see
text); also indicated is half of the equal-mass distance (dotted
lines). The four cases assume a minimum source mass of $1.6 \times
10^8 M_{\odot}$, $1.6 \times 10^9 M_{\odot}$, $1.6 \times 10^{10}
M_{\odot}$, and $1.6 \times 10^{11} M_{\odot}$, respectively from left
to right. Each curve is calculated for a halo of mass equal to the
mass-weighted mean source halo mass (respective values $2.2 \times
10^{10} M_{\odot}$, $4.1 \times 10^{10} M_{\odot}$, $1.1 \times
10^{11} M_{\odot}$, and $4.1 \times 10^{11} M_{\odot}$). All curves
assume $z=6$ and our case of full bias.}
\label{fig-MinM}
\end{figure}

We have thus far adopted the minimum source mass of $1.6 \times 10^8
M_{\odot}$ given by atomic cooling, and have assumed that all halos
above this mass produce ionizing photons with the same overall
efficiency. However, internal stellar feedback effects may be
especially disruptive in source halos with relatively low masses and
low binding energies. Therefore, we consider the possibility of
requiring a higher minimum mass for halos that host efficient ionizing
sources. At the same time we make the individual sources more
efficient, so that the reionization redshift remains approximately
fixed. Figure \ref{fig-MinM} shows that the resulting $f_{\rm esc}$,
for the characteristic source halo mass, increases at a given $R$ as
the minimum source halo mass is increased. The ionizing photons,
however, must reach a more distant $R$ before overlap begins, since
the abundance of ionizing sources is reduced. The final result is that
the screening depends only mildly on the minimum source halo mass. For
the four cases between $1.6\times 10^8 M_\odot$ and $1.6\times 10^{11}
M_\odot$ shown in Figure \ref{fig-MinM}, $f_{\rm esc}=18\%$, $14\%$,
$11\%$, and $9.2\%$, respectively, i.e., $f_{\rm esc}$ is reduced by
only a factor of two when the minimum halo mass is increased by three
orders of magnitude.

Clearly, screening due to fully intact minihalos is a substantial
barrier and can delay the beginning of overlap. Early on, when the
dense gas surrounding each minihalo is still neutral, the screening is
even more effective. For example, when the minimum source mass is
given by atomic cooling, $f_{\rm esc}=18\%$ if we include the gas out
to the virial radius of each minihalo; if we extrapolate the NFW
profile out to twice the virial radius, around each minihalo, the
result is $f_{\rm esc}=4.1\%$. Thus, the surroundings of the minihalos
will be ionized before overlap can make substantial progress, and at
that point the minihalos themselves will begin to
photoevaporate. Figure \ref{fig-MinM} also shows the results for the
later stage discussed in \S \ref{evap}, when only the bound gas
remains in the minihalos. Once the unbound gas has photoevaporated,
the effect of screening no longer substantially impedes the start of
overlap. In particular, the four cases shown in Figure \ref{fig-MinM}
give $f_{\rm esc}=84\%$, $82\%$, $78\%$, and $76\%$, respectively, in
order of increasing minimum halo mass.

The results also depend on additional parameters. For example, our
standard case of $\zeta=0.25$ and $\eta=10\%$ gives $f_{\rm esc}=18\%$
for intact minihalos and $84\%$ for bound minihalos only. Increasing
$\zeta$ to 0.5 gives $38\%$ and $90\%$ for these two cases,
respectively. Note that increasing $\zeta$ at a fixed efficiency makes
each source dimmer but increases the number density of sources, thus
reducing the distance that photons are required to reach in order for
the \ion{H}{2} bubbles of nearby sources to overlap. Reducing $\zeta$
to 0.25 gives $7.1\%$ and $77\%$. The results are less sensitive to
changing $\eta$; setting $\eta=20\%$ gives $24\%$ and $86\%$, while
setting $\eta=5\%$ gives $13\%$ and $83\%$, respectively. Thus, the
detailed numbers depend on the uncertain source parameters, but the
qualitative conclusion remains the same: intact minihalos
substantially impede the overlap stage, while bound minihalos do not.

Although overlap can proceed once only bound minihalos remain,
observationally the universe remains opaque to \Lya radiation. In
other words, the ionizing intensity in the IGM is initially low, and
the small remaining \ion{H}{1} fraction continues to produce a
Gunn-Peterson trough. The trough disappears only when sources can send
their photons out to the rather large distance $R_{\rm GP}$ (see \S
\ref{whichR}). Even the bound minihalos on their own produce (for our
standard parameters, and a typical source) a covering factor of 71 for
total blocking, which implies absolutely total screening for $R=R_{\rm
GP}$. Note that we include in this calculation the effect of the
ionizing background in partially ionizing the minihalos (see \S
\ref{whichR}). The effect of this on the overall screening is small
since in the current context, when a line of sight from a source
encounters a minihalo the typical result is complete blocking even
though the neutral fraction at the virial radius of the minihalo is
$\sim 1\%$; this results from the fact that we are considering only
bound minihalos (which are relatively massive, and thus have high
column densities), and we are considering large distances from the
source (so that the column density or recombination rate per unit
solid angle is high).

Thus, the screening effect can substantially delay the end of the
Gunn-Peterson trough, and this remains true even if the gas in some
minihalos cools due to H$_2$ molecules and only a small fraction of
the minihalos are left intact. The history of molecular cooling in
minihalos prior to reionization involves complicated effects of
radiative feedback, effects which are now beginning to be explored
with a new generation of numerical simulation codes. Although only
redshifts $\ga 10$ can be computed and the results must be considered
tentative, \citet{RGS} found that molecular cooling does occur in some
minihalos, but the total ionizing intensity is limited by a strong
negative feedback of direct ionization of H$_2$ molecules; in
particular, they argued that minihalos cannot contribute substantially
to overlap compared to more massive ionizing sources. Future
simulations may determine the precise fraction of minihalos in which
the gas remains uncooled. We find that even if only $5\%$ of the
minihalos are intact, the bound ones produce essentially complete
blocking (i.e., $f_{\rm esc}=2.5\%$) at $R=R_{\rm GP}$; with $2\%$ of
the minihalos the escape fraction is still small ($23\%$), while $1\%$
of the bound minihalos block around half the photons (i.e., $f_{\rm
esc}=48\%$).

\begin{figure}[htbp]
\plotone{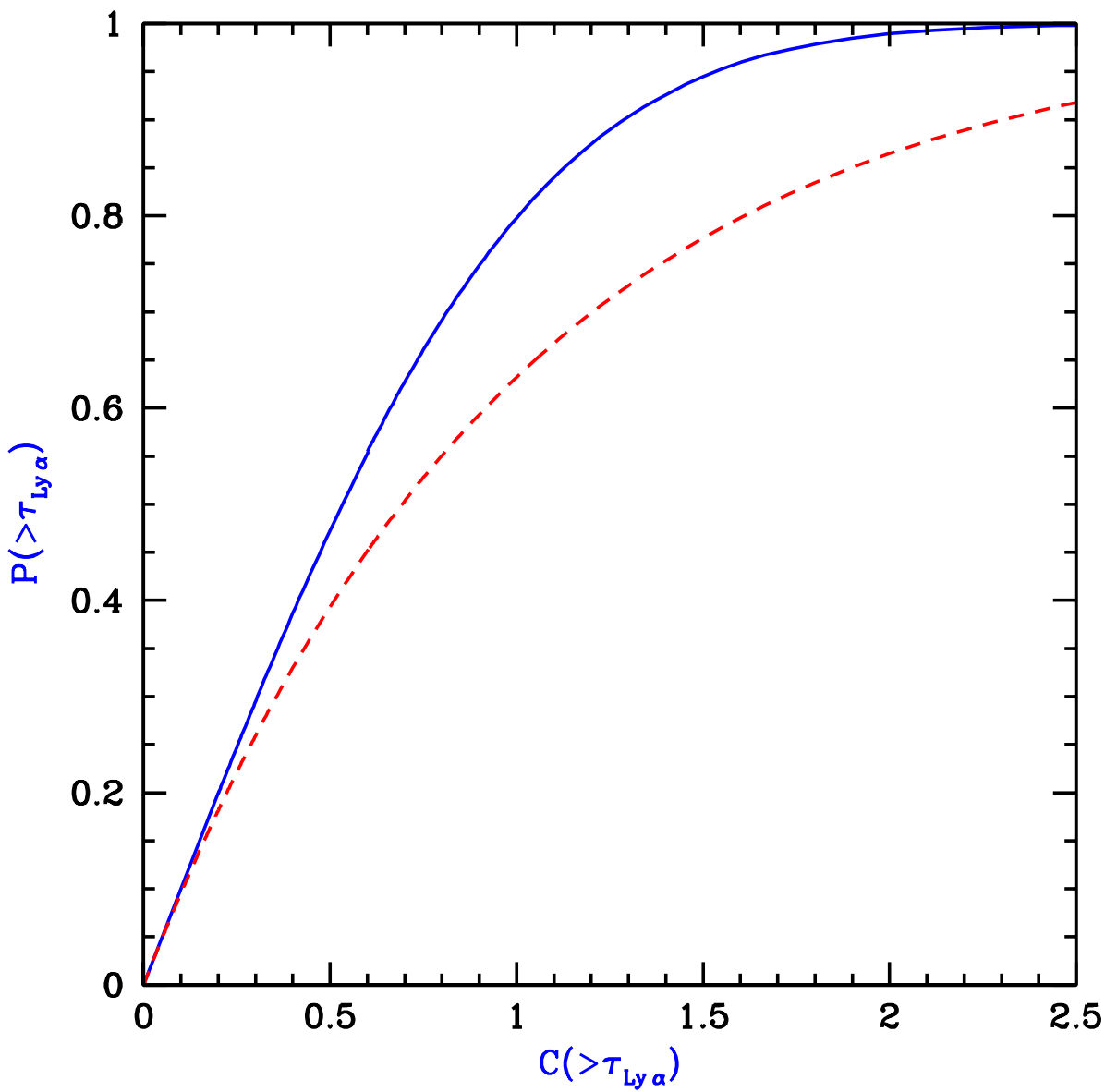} 
\caption{Cumulative probability $P(> \tau_{{\rm Ly}\, \alpha})$ 
for \Lya absorption with optical depth above $\tau_{{\rm Ly}\,
\alpha}$ versus covering factor $C(> \tau_{{\rm Ly}\, \alpha})$. 
Shown are the naive probability derived from the covering factor
(dashed curve), and the actual probability derived with a Monte Carlo
approach (solid curve).}
\label{fig-PLya}
\end{figure}

Finally, we consider the absorption of \Lya photons by minihalos (\S
\ref{Lya}). Since the covering factor (or optical depth) $C$ is a
power law in $\tau_{{\rm Ly}\, \alpha}$, the conversion via Monte
Carlo to a cumulative probability $P(> \tau_{{\rm Ly}\, \alpha})$
involves a universal function, where $P$ is a function of $C$ only and
does not depend separately on $\tau_{{\rm Ly}\, \alpha}$. This
universal function is plotted in Figure \ref{fig-PLya}, where this
function is shown to be significantly different from the naive
probability of $1-e^{-C}$. The probability is $50\%$ when $C=0.53$,
$68\%$ when $C=0.78$, $95\%$ when $C=1.5$, and $99\%$ when $C=2.0$.

For the covering factor itself we obtain with fully intact minihalos,
$C (> \tau_{{\rm Ly}\, \alpha}) = 0.97/\sqrt{\tau_{{\rm Ly}\,
\alpha}}$. Therefore, $P(> 1)=78\%$ of the line of sight is covered
with $\tau_{{\rm Ly}\, \alpha} > 1$ by minihalo damping wings; the
number is $P(> 5)=42\%$ with $\tau_{{\rm Ly}\, \alpha} > 5$. In the
earliest stage, when we can extrapolate the NFW profile out to twice
the virial radius around each minihalo, the corresponding results are
$C (> \tau_{{\rm Ly}\, \alpha}) = 2.3/\sqrt{\tau_{{\rm Ly}\,
\alpha}}$, with $P(> 1)=99.6\%$ and $P(> 5)=81\%$. On the other hand,
in the late stage when only the bound gas remains inside minihalos, $C
(> \tau_{{\rm Ly}\, \alpha}) = 0.083/\sqrt{\tau_{{\rm Ly}\,
\alpha}}$, with $P(> 1)=8.3\%$ and $P(> 5)=3.7\%$. The low optical
depth produced by the bound gas results from the fact that, for a
given total mass, small halos are more effective \Lya absorbers than
large halos. This results from equation (\ref{eq:Lya}): a reduction in
column density by a factor of 2 reduces the distance covered (by a
given $\tau_{{\rm Ly}\, \alpha}$) by only a factor of $\sqrt{2}$. We
conclude that as long as the minihalos are intact, they can produce
significant \Lya absorption, although they are unlikely to themselves
cause full absorption (i.e., $\tau_{{\rm Ly}\, \alpha} > 5$ with a
probability near unity). After the unbound gas photoevaporates,
however, the minihalos can make only a minor contribution to
\Lya absorption.

\section{Conclusions}

In this paper we have shown that the population of gaseous minihalos
can produce substantial screening of ionizing photons during
reionization. After accounting for the biased distribution of
minihalos around sources, we have found that a fully intact population
can delay the start of overlap; for a typical source at redshift
$z=6$, advancing overlap requires getting the ionizing photons out to
a proper distance $R = 73$ kpc, for which the minihalos limit the
escape fraction to a low value of $f_{\rm esc} = 18\%$. The escape
fraction out to the appropriate distance is roughly the same for
source halos over a wide range of mass, with only rare $M > 10^{13}
M_{\odot}$ halos having a much larger escape fraction. Once the
smallest minihalos photo-evaporate, gas remains in the bound minihalos
for a substantially longer period. The bound minihalos are unable to
delay the start of overlap, producing a value of $f_{\rm esc}=84\%$
(for our standard parameters at $z=6$).

To obtain the escape fraction, we have included the effect of
screening due to multiple minihalos (using a Monte Carlo approach),
and also the effect of partial blocking [see equation
(\ref{eq:fesc})]. We typically adopted the minimum source mass of $1.6
\times 10^8 M_{\odot}$ given by atomic cooling; however, if stellar
feedback favors higher-mass halos, we have shown that $f_{\rm esc}$ is
insensitive to the minimum source halo mass if the reionization
redshift is held fixed; and although the detailed numbers depend on
uncertain source parameters (such as efficiency and duty cycle), we
have shown that the overall conclusion remains the same: the full
population of minihalos can substantially impede the overlap stage,
while the bound minihalos themselves cannot.

Even after overlap, screening due to minihalos remains highly
effective until the minihalos photo-evaporate; this screening keeps
the ionizing intensity in the IGM low, and the Gunn-Peterson trough
remains until even the bound minihalos are disrupted. This can produce
a lag of $\sim 200$ Myr (equivalent to redshift interval
$z=7.2\,$--$\,$6) between overlap and the end of Gunn-Peterson
absorption. This screening is so strong that it is highly effective
even if molecular cooling occurs in most minihalos and only a few
percent of them retain uncooled gas.

Intact minihalos may also directly absorb \Lya photons. For the full
population at $z=6$, $42\%$ of a random line of sight is covered with
$\tau_{{\rm Ly}\, \alpha} > 5$. However, as noted above, a full
population of minihalos screens ionizing sources and makes the IGM
absorb \Lya photons much stronger than do the minihalos themselves.
When only the bound minihalos remain, the ionizing intensity is still
low in the IGM, but at this stage the direct absorption due to
minihalos is already insignificant (e.g., only $3.7\%$ of the line of
sight is covered at $\tau_{{\rm Ly}\, \alpha} > 5$).

Current numerical simulations cannot resolve the population of
minihalos during the reionization era. However, we have demonstrated
in this paper that any conclusions regarding the timing of the overlap
era of reionization and of the lifting of Gunn-Peterson absorption
depend crucially on the behavior of the minihalos.

\acknowledgments

This work was supported in part by NASA grants NAG 5-7039, 5-7768, and
NSF grants AST-9900877, AST-0071019 for AL. RB acknowledges the
support of an Alon Fellowship at Tel Aviv University.


\begin{thebibliography}{}

\bibitem[Barkana(2002)]{z6me} Barkana, R. 2002, New Astronomy, 7, 85

\bibitem[Barkana \& Loeb(1999)]{us99} Barkana, R., \& Loeb, A. 1999, 
ApJ, 523, 54

\bibitem[Barkana \& Loeb(2001)]{us01} Barkana, R., \& Loeb, A. 
2001, Phys. Rep., 349, 125

\bibitem[Becker \etalb(2002)]{z6.3} Becker, R. H., Fan, X., White,
R. L., Strauss, M. A., Narayanan, V. K., \etal  2002, AJ, 122, 2850

\bibitem[Bond \etalb(1991)]{B91} Bond, J. R., Cole, S., Efstathiou,
G., \& Kaiser, N. 1991, ApJ, 379, 440

\bibitem[Bullock \etalb(2000)]{cN} Bullock, J. S., Kolatt, T. S., 
Sigad, Y., Somerville, R. S., Kravtsov, A. V., Klypin, A. A., Primack,
J. R., \& Dekel, A. 2000, MNRAS, 321, 559

\bibitem[Cen(2001)]{C01} Cen, R. 2001, ApJ, in press (astro-ph/0101197)

\bibitem[Djorgovski \etalb(2002)]{z5.8} Djorgovski, S. G., Castro,
S. M., Stern, D., \& Mahabal A. 2002, ApJL, 560, 5

\bibitem[Fan \etalb(2002)]{Fan02} Fan, X., Narayanan, V. K., Strauss,
M. A., \etal 2002, AJ, in press (astro-ph/0111184)

\bibitem[Gnedin(2000)]{G00} Gnedin, N. Y. 2000, \apj, 535,
530

\bibitem[Gnedin \& Hui(1998)]{gh98} Gnedin, N. Y., \& Hui, L. 1998, 
MNRAS, 296, 44

\bibitem[Gunn \& Peterson(1965)]{GP} Gunn, J. E., \& Peterson, 
B. A. 1965, ApJ, 142, 1633

\bibitem[Haiman, Thoul, \& Loeb(1996)]{HTL96} Haiman, Z., Thoul,
A. A., \& Loeb, A. 1996, ApJ, 464, 523

\bibitem[Haiman, Abel, \& Madau(2000)]{HAM} Haiman, Z., Abel, T., 
\& Madau, P. 2000, ApJ, 551, 599

\bibitem[Hu \etalb(2002)]{z6.6} Hu, E. M., Cowie, L. L., McMahon, 
R. G., \etal 2002, ApJL, in press (astro-ph/0203091)

\bibitem[Jenkins \etalb(2001)]{jenk00} Jenkins, A., Frenk, C. S.,
White, S. D. M., Colberg, J. M., Cole, S., Evrard, A. E., Couchman,
H. M. P., \& Yoshida, N. 2001, MNRAS, 321, 372

\bibitem[Lacey \& Cole(1993)]{lc93} Lacey, C. G., \& Cole, S. M. 1993,
MNRAS, 262, 627

\bibitem[Lidz \etalb(2002)]{Lidz} Lidz, A., Hui, L., Zaldarriaga, M.,
Scoccimarro, R. 2002, ApJ, in press (astro-ph/0111346)

\bibitem[Loeb \& Rybicki(1999)]{LR99} Loeb, A., \& Rybicki, 
G. B. 1999, \apj, 524, 527

\bibitem[Miralda-Escud\'e(1998)]{jordi98} Miralda-Escud\'e, J. 1998, 
ApJ, 501, 15

\bibitem[Navarro, Frenk, \& White(1997)]{NFW} Navarro, J. F., 
Frenk, C. S., \& White, S. D. M. 1997, ApJ, 490, 493 (NFW)

\bibitem[Press \& Schechter(1974)]{PS74} Press, W. H., 
\& Schechter, P. 1974, ApJ, 187, 425

\bibitem[Ricotti, Gnedin, \& Shull(2002)]{RGS} Ricotti, M., Gnedin,
N. Y., \& Shull, J. M. 2002, ApJ, in press

\bibitem[Razoumov \etalb(2002)]{R02} Razoumov, A. O., Norman, M. L., 
Abel, T., \& Scott, D. 2002, ApJ, in press; astro-ph/0109111

\bibitem[Scalo(1998)]{scalo} Scalo, J. 1998, in ASP conference series
Vol 142, The Stellar Initial Mass Function, eds. G. Gilmore \&
D. Howell, p. 201 (San Francisco: ASP)

\bibitem[Scannapieco \& Barkana(2002)]{mePast} 
Scannapieco, E., \& Barkana, R. 2002, ApJ, in press

\bibitem[Sheth \& Tormen(1998)]{st98} Sheth, R. K., \& Tormen,
G. 1998, MNRAS, 300, 1057

\bibitem[Thoul \& Weinberg(1996)]{TW} Thoul, A. A., \& Weinberg, 
D. H. 1996, ApJ, 465, 608

\end{thebibliography}
\end{document}